\documentclass[runningheads]{llncs}
\usepackage{marvosym}
\usepackage[T1]{fontenc}
\usepackage{graphicx,verbatim}
\usepackage{amsmath,amssymb}
\usepackage{multicol,multirow}
\usepackage{booktabs}
\usepackage{hyperref}
\begin{document}
\title{LesionDiffusion: Towards Text-controlled General Lesion Synthesis}
%

\author{Wenhui Lei\inst{1,2,\dagger} \and Hengrui Tian\inst{1,\dagger} \and  Linrui Dai\inst{1,2\dagger} \and Hanyu Chen\inst{3} \and Xiaofan Zhang\inst{1,2,}$^{\textrm{\Letter}}$}

\authorrunning{H. Tian and W. Lei et al.}
\institute{Shanghai Jiaotong University \and Shanghai Artificial Intelligence Laboratory  \and The First Hospital of China Medical University \\
    \email{xiaofan.zhang@sjtu.edu.cn}}
    
\renewcommand{\thefootnote}{\fnsymbol{footnote}}
\footnotetext{$\dagger$ Contributed equally to this work.}
    
\maketitle              
\begin{abstract}
Fully-supervised lesion recognition methods in medical imaging face challenges due to the reliance on large annotated datasets, which are expensive and difficult to collect. To address this, synthetic lesion generation has become a promising approach. However, existing models struggle with scalability, fine-grained control over lesion attributes, and the generation of complex structures. We propose LesionDiffusion, a text-controllable lesion synthesis framework for 3D CT imaging that generates both lesions and corresponding masks. By utilizing a structured lesion report template, our model provides greater control over lesion attributes and supports a wider variety of lesion types. We introduce a dataset of 1,505 annotated CT scans with paired lesion masks and structured reports, covering 14 lesion types across 8 organs. LesionDiffusion consists of two components: a lesion mask synthesis network (LMNet) and a lesion inpainting network (LINet), both guided by lesion attributes and image features. Extensive experiments demonstrate that LesionDiffusion significantly improves segmentation performance, with strong generalization to unseen lesion types and organs, outperforming current state-of-the-art models. Code is available at \href{https://github.com/HengruiTianSJTU/LesionDiffusion}{here}.



\end{abstract}
\section{Introduction}
Deep learning has driven significant advances in medical image analysis, especially in lesion detection and diagnosis \cite{jiang2024unleashing,lei2025dataefficientpantumorfoundationmodel,lei2024medlsam,lei2023one}. However, current fully-supervised lesion recognition methods rely heavily on large annotated datasets, which are costly to collect and share due to privacy concerns.

To address the scarcity of lesion data, synthetic lesion generation has emerged as a promising solution. In 3D CT imaging, techniques such as GANs \cite{goodfellow2020generative}, diffusion models \cite{sohl2015deep,ho2020denoising}, and physical simulations have been used to generate synthetic lesions for various conditions\cite{huang2025interactive,yu2024ct}, including lung nodules \cite{jin2018ct}, COVID-19 \cite{yao2021label}, and liver tumors \cite{hu2023label}. However, these methods train each lesion type independently, which limits their scalability and generalizability. Additionally, they lack fine-grained control over lesion attributes, and are primarily designed to simulate solid lesions, neglecting complex structures such as hollow lesions found in colorectal or gastric cancers \cite{chen2024towards}. This makes them hard to serve as a general-purpose, controllable 3D CT lesion synthesis model.


In this work, we aim to develop a general, text-controllable lesion synthesis model for 3D CT imaging that generates both lesions and their corresponding masks. Our approach addresses the limitations of previous lesion synthesis methods, which include: 1) the limited variety of lesion types, primarily due to the scarcity of publicly available CT datasets that include both lesion descriptions and corresponding masks; 2) insufficient control over lesion attributes; and 3) the challenges in generating complex lesion textures and masks.

\begin{figure*}[t!]
    \centering
    \includegraphics[width=1\textwidth]{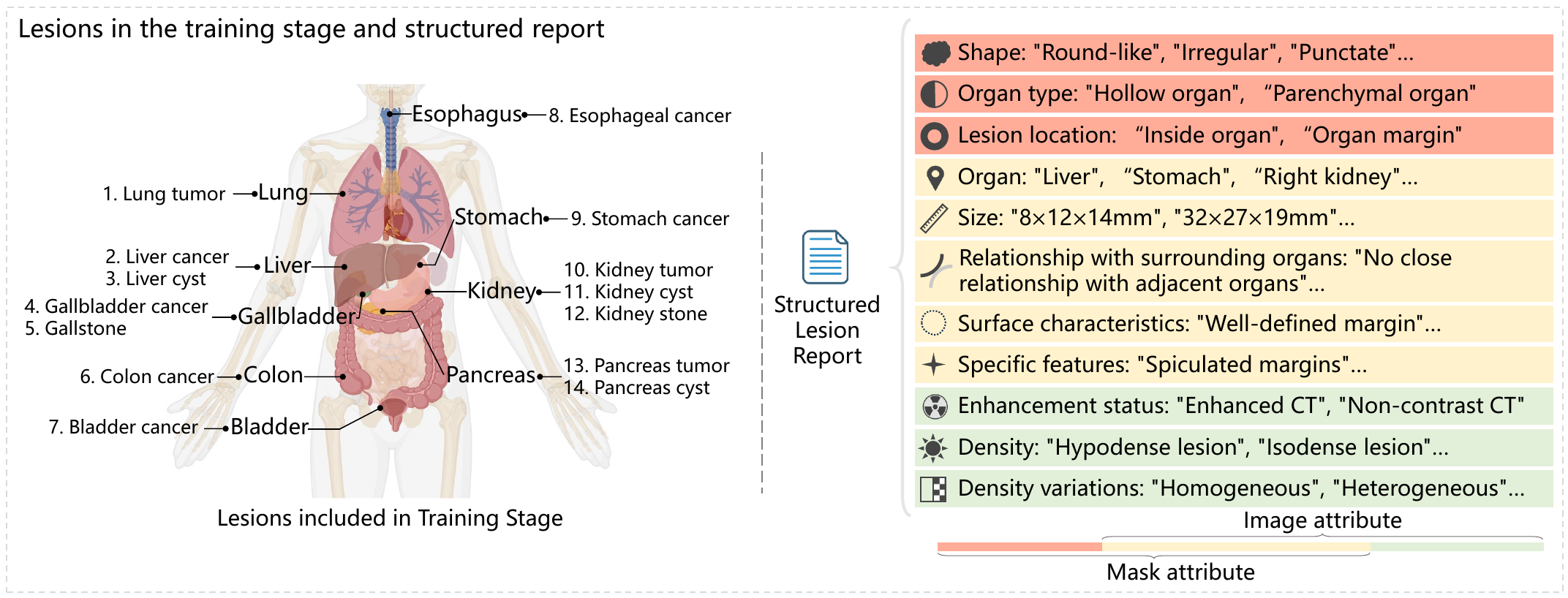}
    \caption{\textbf{Lesions and Structured Lesion Reports in the Training Stage.} The LesionDiffusion model is trained on 14 types of lesions across 8 organs. The corresponding structured reports include 10 attributes, which are categorized into mask attributes and image attributes.
}
    \label{fig:attribute}
\end{figure*}

\begin{figure*}[t!]
    \centering
    \includegraphics[width=1\textwidth]{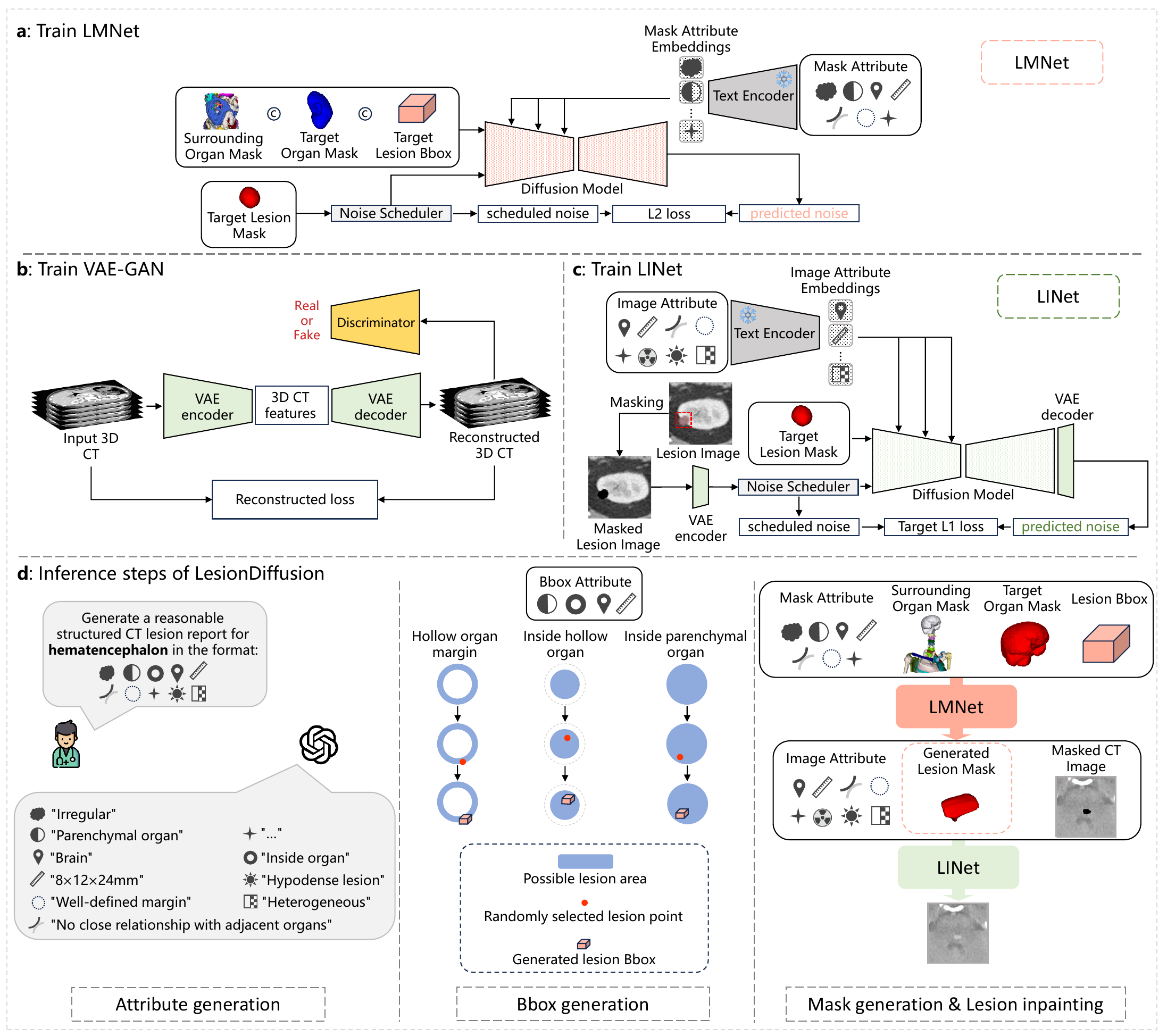}
    \caption{\textbf{Overview of LesionDiffusion framework.} (a) LMNet is trained to generate lesion masks; (b) VQ-GAN is trained to compress 3D CT images into latent space and then reconstruct them; (c) LINet is trained to perform lesion inpainting in the latent space; (d) During the inference stage, the framework generalizes to any lesion type, involving lesion attribute generation, lesion bounding box generation, lesion mask generation, and inpainting.
}
    \label{fig:overview}
\end{figure*}

To overcome these limitations and enhance the generalizability and controllability of lesion synthesis, we adopt a structured lesion report template, as proposed in \cite{lei2025dataefficientpantumorfoundationmodel}, to serve as the textual condition. This template, which covers various aspects of lesion attributes (as illustrated in Fig. \ref{fig:attribute}), enables precise control over lesion mask generation and corresponding image inpainting based on specified attributes. Additionally, we have collected and annotated 1,505 CT scans with paired lesion masks and structured reports, covering 14 types of lesions across 8 organs. Building on this dataset, we developed \textbf{LesionDiffusion} (Fig. \ref{fig:overview}), a text-controlled framework for generalizable lesion synthesis. The framework consists of two key components: 1) a lesion mask synthesis network (LMNet) that is guided by lesion bounding boxes (bbox) and mask attributes, and 2) a lesion inpainting network (LINet) that is guided by both image attributes and the lesion mask. Through extensive experiments, we demonstrate that LesionDiffusion significantly improves segmentation performance across a wide range of lesion types. More importantly, our approach shows exceptional generalization, even for unseen organs and lesion types, achieving a notable improvement in lesion segmentation and surpassing the performance of existing state-of-the-art lesion synthesis models.

\section{Methods}

\subsection{Stage I: Lesion Mask Generation}
Concurrent tumor inpainting methods, like DiffTumor~\cite{chen2024towards}, adopt an end-to-end patch-wise generation scheme of tumor CTs based on a tumor mask deformed from a unit sphere. Although this generation scheme is plausible in organs whose tumors are characterized by a round-like shape and an arbitrary position, it undermines generalizability as more tumor types do not share these features. 




To effectively translate vague shape information into concrete lesion masks, mask-related attributes from structured reports—such as shape, organ type, and location—are used as conditional guidance, as illustrated in Fig.\ref{fig:attribute} and Fig.\ref{fig:overview} (a). 

Next, we develop a diffusion model, named LMNet, to generate lesion masks that adhere to these constraints. Formally, for a precise lesion annotation $\mathbf{m}_0$, the forward noising process can be described as 
\begin{equation}
\mathbf{m}_t = \sqrt{\bar{\alpha}_t} \mathbf{m}_0 + \sqrt{1 - \bar{\alpha}_t}\boldsymbol\epsilon,
\label{eq:noise_addition}
\end{equation}
where $\mathbf{m}_t$ is the noisy mask at timestep $t$ and $\boldsymbol{\epsilon}$ is the Gaussian noise, $\bar{\alpha}_t$ represents the noise variance schedule~\cite{ho2020denoising}. Denoting the annotated mask attributes as $\mathcal{A}_{\rm mask}$, the $\mathbf{m}_0$ surrounded bounding box as $\mathcal{B}$ and the target organ semantics as $\mathcal{M}$, we proposed a bbox-weighted loss function with a hyperparameter $\gamma$:
\begin{equation}
    \begin{split}
    \mathcal{L}_{\rm LMNet} &=  \mathbb{E}_{\boldsymbol\epsilon \sim \mathcal{N}(0, I)} \left[ \| \boldsymbol\epsilon \odot \mathcal{B} - \boldsymbol\epsilon_\theta(\mathbf{m}_t; t, \mathcal{M}, \mathcal{B}, \mathcal{A}_\text{mask}) \odot \mathcal{B} \|_2^2 \right. \\
    & \left. + \gamma \| \boldsymbol\epsilon - \boldsymbol\epsilon_\theta(\mathbf{m}_t; t, \mathcal{M}, \mathcal{B}, \mathcal{A}_\text{mask}) \|_2^2 \right]
    \end{split}
    \label{eq:weighted_loss}
\end{equation}
This objective is proposed to encourage the model's focus on the constrained area specified by our bounding box $\mathcal{B}$, thus promoting accurate synthesis of the lesion masks. The target organ semantic is extracted from the surrounding organ semantics as a separate channel before concatenating with the bounding box $\mathcal{B}$ and serving as a spatial condition for tumor site generation. Mask attributes $\mathcal{A}_{\rm mask}$ are embedded using a pretrained text encoder in BiomedCLIP~\cite{zhang2025biomedclipmultimodalbiomedicalfoundation} and fused into the generation process via cross-attention modules to describe shape, location and other pertinent spatial information to the diffusion model.

\subsection{Stage II: Lesion Image Inpainting}
Similar to the mask-attributes-guided generation of the lesion mask, we utilize image attributes—such as density, variation, and surface characteristics—to perform lesion image inpainting in Stage II, as illustrated in Fig.\ref{fig:attribute} and Fig.~\ref{fig:overview} (c).

Due to the large memory consumption of 3D CT images, we choose  LatentDiffusion~\cite{rombach2022high} to train the diffusion model on a smaller latent space encoded by a pair of dedicated autoencoders $\mathcal{E}_{\rm CT}$, $\mathcal{D}_{\rm CT}$.
This is done following a similar paradigm as VQ-VAE~\cite{van2017neural}, which encodes images with a discretized codebook $\mathbf{c}$ and quantization step $\mathbf{q}_\mathbf{c}$. We also adopt an adversarial module $\mathcal{D}_{\rm disc}$ to boost autoencoding performance similar to DiffTumor~\cite{chen2024towards}. For each CT image $\mathbf{x}$, denoting its encoded latent as $\mathbf{z}=\mathcal{E}_{\rm CT}(\mathbf{x})$ and its reconstructed counterpart as $\hat{\mathbf{x}}=\mathcal{D}_{\rm CT}(\mathbf{q}_\mathbf{c}(\mathbf{z}))$, the training objective of our autoencoder is defined as the combination of $l_2$ reconstruction loss, codebook loss~\cite{esser2021taming} and discriminative loss from the adversarial module 
\begin{equation}
\begin{split}
        \mathcal{L}_{\rm AE}&=||\mathbf{x}-\hat{\mathbf{x}}||^2_2+||\Omega(\mathbf{z})-\mathbf{q}_\mathbf{c}(\mathbf{z})||_2^2+||\Omega(\mathbf{q}_\mathbf{c}(\mathbf{z}))-\mathbf{z}||_2^2\\&+\log\mathcal{D}_{disc}(\mathbf{x})+\log(1-\mathcal{D}_{disc}(\hat{\mathbf{x}}))\\
\end{split}
\end{equation}
where $\Omega$ is the stop-gradient function. After training the paired autoencoder, we develop another diffusion model, dubbed LINet, on the latent space of $\mathcal{E}_{\rm CT}$. However, since we focus only on a specific lesion region guided by the annotated lesion masks $\mathbf{m}$, we choose to perform the diffusion process solely within the lesion site. Formally, the forward diffusion process can be described as
\begin{equation}
    \mathbf{z}_t=(\sqrt{\bar{\alpha}_t}\mathbf{z}_0+\sqrt{1-\bar{\alpha}_t}\boldsymbol{\epsilon})\odot{\rm Resize}(\mathbf{m})+\mathbf{z}_0\odot(1-{\rm Resize}(\mathbf{m}))
\end{equation}
where ${\rm Resize}(\cdot)$ function is used to resize the semantic condition $\mathbf{m}$ to be of the same spatial resolution as the image latent $\mathbf{z}_0$, $\boldsymbol{\epsilon}$ is the Gaussian noise, and $\bar{\alpha}_t$ represents the noise variance schedule. The denoising step uses LINet to retrieve inpainted image latent from noises conditioned on image attributes $\mathcal{A}_{\rm image}$ encoded by BiomedCLIP.
To optimize the performance of the diffusion model, we leverage the bounding box \(\mathcal{B}'\) surrounding the lesion area in the mask and use $l_1$ loss to measure the difference between the input noise \(\boldsymbol\epsilon\) and the denoised latents \(\boldsymbol\epsilon_\theta\) within this box. Formally, the training objective is 
\begin{equation}
    \begin{aligned}
    \mathcal{L}_{\rm LINet} = \mathbb{E}_{\epsilon \sim \mathcal{N}(0, I)} \left[ \frac{1}{|\mathcal{B}'|} \| \boldsymbol\epsilon \odot \mathcal{B}' - \boldsymbol\epsilon_{\theta}(\mathbf{z}_t; t, {\rm Resize}(\mathbf{m}), \mathcal{A}_{\rm image}) \odot \mathcal{B}' \|_1 \right]
    \end{aligned}
\label{eq:bbox_loss}
\end{equation}

\subsection{Inference process}
In the inference stage of LesionDiffusion, given the lesion to be simulated, we first generate its structured report and then sample to obtain the bounding box (bbox). Finally, we use LMNet and LINet to obtain the lesion mask and inpainted CT images, as shown in Fig. \ref{fig:overview}.

Specifically, we propose two methods for generating the complete structured lesion report: 1) the user selects possible options for each attribute of the target lesion based on existing medical knowledge, followed by random sampling and combination; 2) the user queries a large language model (LLM) to construct a reasonable report. Once the report is obtained, we extract the bbox attributes: "organtype," "lesion location," "organ," and "size." Based on these attributes, we categorize the potential locations of the lesion into: "hollow organ margin," "inside hollow organ," and "inside parenchymal organ," and randomly sample within the corresponding regions to generate the bbox.

Next, the lesion mask attribute, surrounding organ mask, target organ mask, and lesion bbox are fed into LMNet. We use DDIM~\cite{song2020denoising} for fast sampling from the learned lesion mask distribution. Since the generation result is a continuous floating-point intensity map, we apply a simple capping threshold to obtain a binary tumor semantic mask. For LINet inference, firstly VQ-GAN encoder compresses CT image input into latent space. Then Gaussian noise replaces the latent codes within the mask region, resulting in \(z_T = \epsilon \odot m + z_0 \odot (1 - m)\), and execute \(T\) time steps of reverse denoising to obtain the edited latent code, which is then decoded back into a CT image by the VQ-GAN decoder.

\section{Experiment}
\paragraph*{Dataset Construction}
We compiled several datasets for generative training and downstream evaluation. Specifically, we assemble 1505 CT scans and masks collected from public and private sources: KiTS23~\cite{heller2023kits21}: kidney tumor and cyst (489 scans); MSD~\cite{antonelli2022medical}: colon tumor (126 scans), liver tumor (303 scans), lung tumor (96 scans), pancreas tumor (216 scans), pancreas cyst (65 scans); private data collected at The First Hospital of China Medical University: liver cyst (30 scans), gallbladder cancer (30 scans), gallstones (30 scans), esophageal cancer (30 scans), gastric cancer (30 scans), kidney stone (30 scans), bladder cancer (30 scans). All these scans are annotated with structured lesion reports by four radiologists. The dataset was partitioned into training and testing sets with a 4:1 ratio for each lesion, resulting in 1204 training samples and 301 testing samples. We also set a hematencephalon segmentation dataset, INSTANCE22~\cite{li2023stateoftheart3danisotropicintracranial} (100 scans, train:test=80:20), for the generalization test, which is not included in the generative model training set. We further collected 10767 CT images at The First Hospital of China Medical University as templates for the inpainting stage. 

\paragraph*{Implementation Details}
All generative trainings are performed under a static learning rate of $1\times 10^{-4}$ with a batch size of 1 per GPU and a patch size of $128\times128\times128$ on 4 NVIDIA 4090 GPUs. The training processes use AdamW~\cite{loshchilov2017decoupled} optimizers with a momentum of 0.99 and a weight decay of $1\times10^{-5}$. The diffusion model uses cosine noise schedule with 1000 timesteps and DDIM denoising~\cite{song2020denoising} of 200 timesteps to recover CT latents from Gaussian noise at a guidance scale of 2.5. Moreover, our image autoencoder downsamples the original image by $2\times$ and uses a codebook of 2048 entries. $\gamma$ is set to 0.001. For the downstream segmentation tasks, we train each segmentation network using the official implementation of nnUNet-v2. Performance was evaluated for each lesion under three settings: (1) training the segmentation model using 1000 synthetic lesions generated from template data; (2) fine-tuning the nnUNet-V2 model, initially trained on synthetic data, with real data from the corresponding lesion's training set; (3) training the segmentation model exclusively on the real dataset. The same settings were used when evaluating other synthesis models.


\paragraph*{Downstream usability}
\begin{table*}[t] 
    \centering
        \caption{DSC comparison under three settings real lesion data only, synthetic lesion data only, and synthetic lesion data augmented by real lesion data.}
    \resizebox{\linewidth}{!}{
    \begin{tabular}{l|cc|ccc|ccccccccc|c}
    \toprule
        \multirow{2}{*}{Methods}&\multicolumn{2}{c|}{Stone}&\multicolumn{3}{c|}{Cyst}&\multicolumn{9}{c|}{Cancer}&\multirow{2}{*}{Avg.}\\
        &Gall.&Kid.&Liver&Pan.&Kid.&Eso.&Gall.&Lung&Liver&Pan.&Kid.&Sto.&Colon&Blad.&\\\midrule
        Real&77.6 &70.4 &76.4 &78.2 &59.8 &57.7 &60.4 &69.0 &61.8 &49.6 &79.2 &66.0 &58.1 &83.3 &67.7
        \\\midrule
        DiffTumor~\cite{chen2024towards}&58.1 &17.5 & 61.6 &66.2 &55.1 &44.8 &49.2 &54.8 &44.6 &30.1 &59.7 &56.4 &41.6 &65.2& 50.4
\\
DiffTumor+Real&75.1&54.4&84.4&79.2&63.1&62.0&63.4&66.7&48.0&44.1&81.9&71.3&59.9&79.5&66.6\\
LesionDiffusion&71.9&21.0&77.2&65.7&48.3&60.4&55.7&61.3&63.8&35.8&60.5&61.1&42.6&76.2&57.3\\
        LesionDiffusion+Real&77.2 &67.0 &83.0 &80.4 &62.5 &61.8 &66.0 &68.9 &64.7 &51.7 &82.6 &70.0 &61.4 &84.2 &70.1 \\
        \bottomrule
    \end{tabular}
    }
    \label{tab:main}
\end{table*}

We first validate LesionDiffusion's competence as a text-based dataset augmentator by comparing the performance of dedicated segmentation models trained with and without LesionDiffusion-synthesized samples. For reference, we also compare LesionDiffusion's augmentation performance with another lesion-generating framework, DiffTumor. As shown in Tab.~\ref{tab:main}, samples synthesized by LesionDiffusion can provide segmentation guidance comparable to that of real samples and far better than that of DiffTumor-synthesized samples. Specifically, we observe an averaged 6.9\% DSC increase in segmentation models pre-trained with LesionDiffusion-synthesized lesions instead of DiffTumor. These models also performs the best on average after task-specific finetuning, outperforming models trained only with real data by 2.4\% and models pre-trained with DiffTumor-synthesized lesions by 3.5\%.

\begin{figure*}[t]
    \centering 
    \includegraphics[width=1\textwidth]{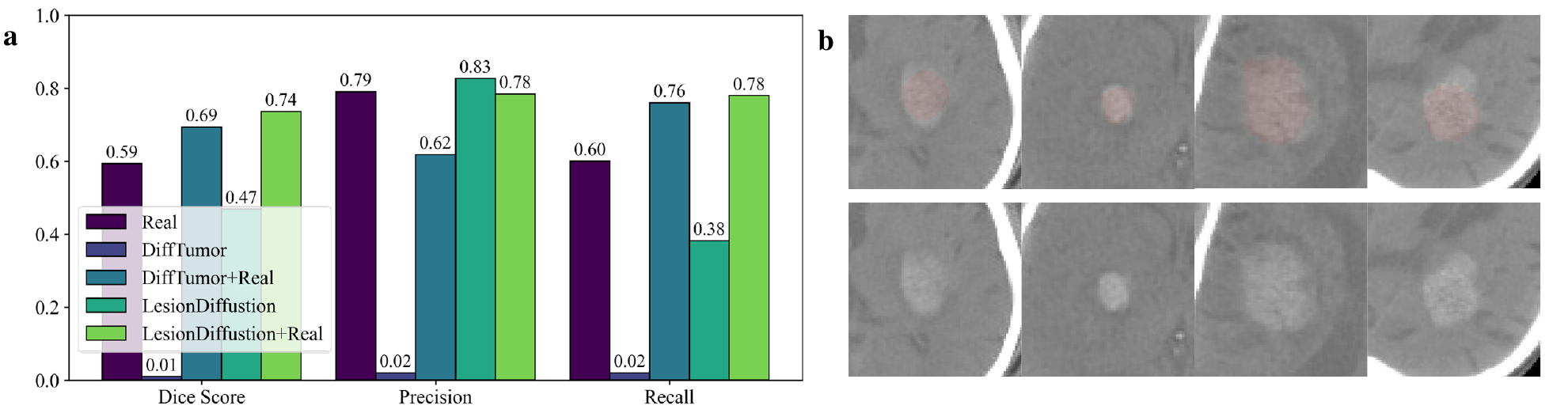}
    \caption{\textbf{Generalization to hematencephalon.} (a) Downstream segmentation results for hematencephalon. (b) Examples synthesized by LesionDiffusion.
}
    \label{fig:domain}
\end{figure*}

\paragraph*{Downstream Generalization}
We also expect the synthetic ability of LesionDiffusion to be generalizable across both seen and unseen downstream tasks. Thanks to our comprehensive structured medical report template which encompasses key attributes for describing lesion shape and texture, LesionDiffusion can be easily transferred to new tasks by using only an appropriate set of attribute inputs, as shown in Fig.~\ref{fig:domain}.
To quantify LesionDiffusion's generalization ability, we use it to synthesize samples of hematencephalon for supervising a dedicated segmentation model. As illustrated in Fig.~\ref{fig:domain}(a), despite LesionDiffusion's original training set lacking brain CT images and hemorrhage lesion types, models trained on LesionDiffusion synthesized samples can still provide segmentation guidance comparable to that of real samples. In contrast, DiffTumor suffers significantly from this domain gap, with a 46\% drop in DSC compared to LesionDiffusion. Moreover, after task-specific finetuning, model trained using LesionDiffusion synthesized lesions manages to provide a segmentation performance 15\% better than that trained with real data and 5\% better than that trained with DiffTumor. 

\begin{figure*}[h]
    \centering
    \includegraphics[width=\linewidth]{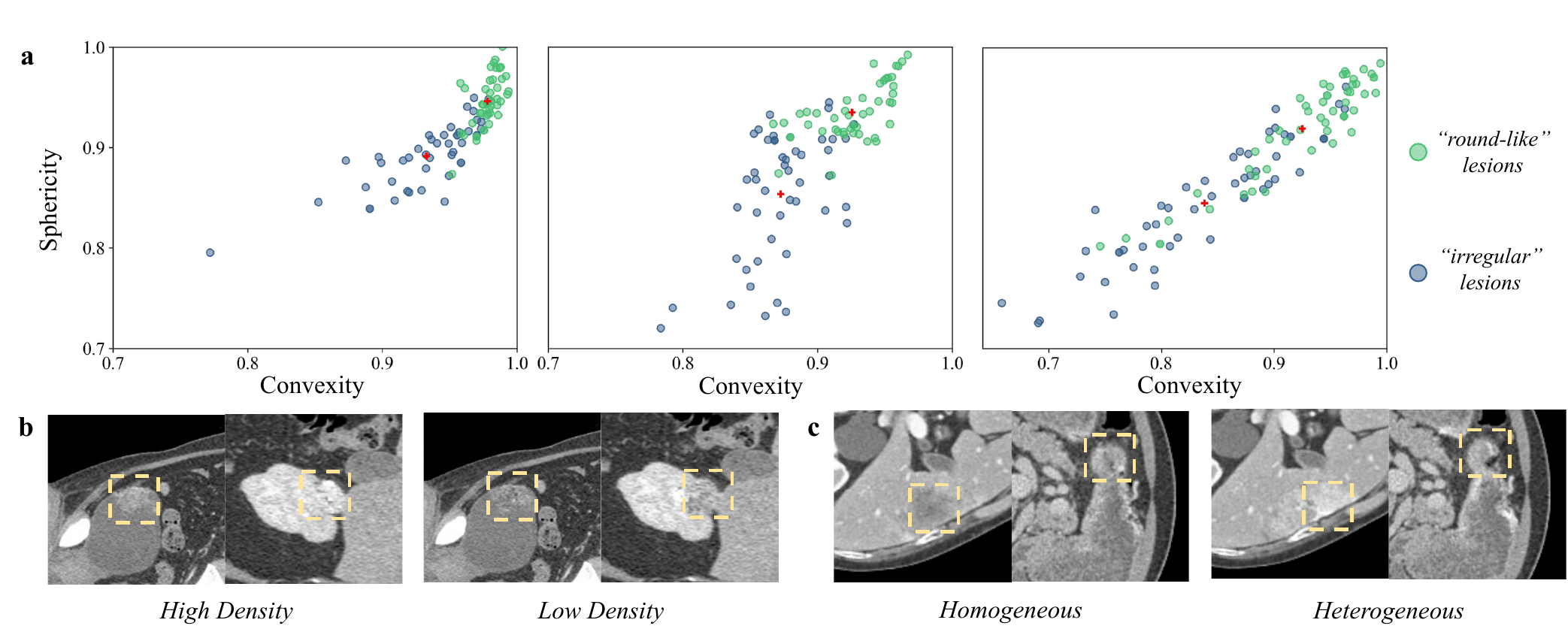}
    \caption{\textbf{Generation alignment with textual guidance.} This figure illustrates the effect of shape, density, and density variation conditions on image inpainting results. (a) Green and blue dots represent the morphological metrics computed for round-like and irregular tumor volumes respectively, with red crosses indicating the metric mean values. (b) Comparison of generation control under the condition of density. (c) Comparison of generation control under the condition of density variation.}
    \label{fig:geo}
\end{figure*}

\paragraph*{Generation alignment with textual guidance}
Firstly, 
we quantitatively evaluate LMNet's conditional alignment between generated tumor shape and prompted attributes using two morphological metrics individually evaluating the convexity~\cite{6247666} and sphericity~\cite{s24030955} of generated masks. Mathematically, a higher convexity and sphericity indicate that the lesion more closely resembles a spherical shape. As shown in Fig.~\ref{fig:geo}(a), we can see a clear trend that masks generated with "round-like" prompts are higher in both metrics compared to those with "irregular" prompts. Additionally, we qualitatively evaluate LINet's conditional alignment by alternating several image attributes in Fig.~\ref{fig:geo}(b,c), the results show promising conformity with the input conditions.


\paragraph*{Ablation study of textual condition} Ablation studies of LesionDiffusion are performed by removing textual conditions in the generation process, forcing the model to only implicitly learn lesion features through the training process. As shown in Tab.~\ref{tab:ablation}, the results show different degrees of decrease in downstream performances with respect to all different tasks, with an averaged decrease of 15.4\% DSC. This effect is particularly pronounced for organs containing multiple lesion types in the training set, such as the liver, kidney, and pancreas, and is primarily due to the absence of textual conditions, which causes the model to be uncertain about the specific lesions to generate in these organs.
\begin{table*}[t] 
    \centering
    \caption{Ablation studies of LesionDiffusion.}
    \resizebox{\linewidth}{!}{
    \begin{tabular}{l|cc|ccc|ccccccccc|c}
    \toprule
        \multirow{2}{*}{Settings}&\multicolumn{2}{c|}{Stone}&\multicolumn{3}{c|}{Cyst}&\multicolumn{9}{c|}{Cancer}&\multirow{2}{*}{Avg.}\\
        &Gall.&Kid.&Liver&Pan.&Kid.&Eso.&Gall.&Lung&Liver&Pan.&Kid.&Sto.&Colon&Blad.&\\\midrule
LesionDiffusion&71.9&21.0&77.2&65.7&48.3&60.4&55.7&61.3&63.8&35.8&60.5&61.1&42.6&76.2&57.3\\
        LesionDiffusion $w.o.$ text&62.4&15.1&42.8&48.6&24.9&60.2&50.7&57.5&44.3&20.1&41.9&33.7&31.8&53.1&41.9 \\
        \bottomrule
    \end{tabular}
    }

    \label{tab:ablation}
\end{table*}

\section{Conclusion}
In conclusion, we introduce LesionDiffusion, a text-controllable framework for inpainting lesions in 3D CT images. By utilizing a structured lesion report template, the framework enables fine-grained control over lesion attributes and accommodates a diverse range of lesions. Our model demonstrates substantial improvements in segmentation performance and exhibits strong generalization to unseen lesions and organs. LesionDiffusion outperforms current state-of-the-art models, providing a scalable solution for lesion recognition in medical imaging. While promising, further research is needed to explore its generalization across a broader range of diseases, and the training dataset is still limited. In future work, we aim to expand the training dataset and validation scenarios to develop a controllable lesion generation foundation model.

\begin{credits}
\subsubsection{\ackname} This work was supported by the National Natural Science Foundation of China (82203199)(H.C.), (62301311)(X.Z.).

\end{credits}

%
%
%
\bibliographystyle{splncs04}
\bibliography{bib}
%




\end{document}